\documentclass[12pt]{article}

\usepackage[totalheight = 23cm, totalwidth = 17cm]{geometry}
\usepackage{amssymb,amsmath,amsfonts,amsbsy,graphicx,xcolor}

\newcommand{\calL}{{\cal L}}
\newcommand{\calO}{{\cal O}}

\usepackage{color}

\begin{document}

\begin{titlepage}

\begin{center}

\rightline{APCTP-Pre2014-006}

\vskip 1cm

\Huge{Inflation beyond T-models and primordial $B$-modes}

\vskip 1cm

\large{
Yi-Fu Cai$^a$\footnote{E-mail address: yifucai@physics.mcgill.ca.},
\hspace{0.1cm}
Jinn-Ouk Gong$^{b,c}$\footnote{E-mail address: jinn-ouk.gong@apctp.org.},
\hspace{0.1cm} and \hspace{0.1cm}
Shi Pi$^b$\footnote{E-mail address: spi@apctp.org.}
}

\vskip 1.2cm

\small{\it
$^{a}$Department of Physics, McGill University, Montr\'eal, QC H3A 2T8, Canada
\\
$^{b}$Asia Pacific Center for Theoretical Physics, Pohang 790-784, Korea
\\
$^{c}$Department of Physics, Postech, Pohang 790-784, Korea
}

\vskip 1.2cm

\end{center}

\begin{abstract}

We describe extended 
theories which shares the gauge transformation symmetry of the T-models, and takes the T-models as well as Starobinsky model as special cases. We derive a general relation between the two slow-roll parameters, and find that a large class of models can be embedded. Such models include more general Starobinsky-like inflation as well as the chaotic inflation with a large tensor-to-scalar ratio consistent with the BICEP2 result.

\end{abstract}

\end{titlepage}

\setcounter{page}{0}
\newpage
\setcounter{page}{1}

\section{Introduction}
\label{sec:intro}

Inflation, as proposed in early 1980s to address conceptual issues of the hot big bang cosmology~\cite{inflation}, has become the most successful paradigm of describing the very early universe. Its profound prediction of a nearly scale-invariant primordial power spectrum has been verified to high precision by the observations on the cosmic microwave background (CMB) in recent years~\cite{Ade:2013zuv}. Moreover, recently the BICEP2 collaboration reported the detection of the CMB primordial $B$-mode polarization~\cite{Ade:2014xna}. This observation, if confirmed to be originated from primordial gravitational waves, will be another major significance for inflation.

The success of inflation, however, is based on a series of assumptions including the existence of a sufficiently long period of quasi-exponential expansion realized by an as yet undetermined physical mechanism, such as to introduce a slow-roll scalar field or to modify gravity theories. An important question to be addressed is how to formulate the effective field description from fundamental theories. In the literature, there have been many attempts of realizing the (generalized) Starobinsky model~\cite{Starobinsky:1980te} in the context of, namely, the supergravity descriptions~\cite{sugra-starobinsky}, the approach of asymptotically safe gravity~\cite{safe-starobinsky} and the modified gravitational actions~\cite{modifiedgra-starobinsky}.

Recently, a 
 new class of models was developed in~\cite{Ferrara:2010in}, which suggested that inflation models with the same conformally described gauge invariance are observationally equivalent. Specifically, a gravitational system involving a scalar field with a flattened potential can be derived by 
two-field theory with an approximate $SO(1,1)$ symmetry~\cite{Kallosh:2013lkr} which is invariant under a gauge transformation. This class of models was dubbed as T-models later~\cite{Kallosh:2013hoa}. The T-models, which can be realized in supergravity, predict small tensor-to-scalar ratio similar to Starobinsky model. This original version of the T-models is however under danger because of the large $B$-mode detected by BICEP2~\cite{Ade:2014xna}. In this paper we study possible extensions of T-models which are invariant under this same gauge transformation. We investigate the properties of the T-models and find that they can be embedded into a broader class which includes specific typical examples that covers a larger portion on the $r$-$n_s$ plane. Especially, taking the recent observation on large $B$-modes in consideration, we can seek for a simple chaotic inflation model with a power-law potential~\cite{Linde:1983gd} in such a class of conformally described models.

This article is organized as follows. In Section~\ref{sec:model} we briefly review the T-models and the gauge transformation under which the Lagriangian is invariant. 
Then, we extend it to a generalized description by introducing a non-trivial potential beyond the T-models. In Section~\ref{sec:general} we focus on some specific examples and study their cosmological implications, especially the predictions on the $r$-$n_s$ relationship. We compare the predictions with
the recent BICEP2 data on tensor modes. We then conclude in Section~\ref{sec:conc}.

\section{Conformally invariant model}
\label{sec:model}

Let us start by quickly reviewing the T-models 
proposed in~\cite{Kallosh:2013lkr}. If we write down a non-minimally coupled Lagrangian with two real scalar fields,
\begin{equation}\label{conformalStar}
\calL = \sqrt{-g}
\left[ \frac{R}{12} \left( \chi^2-\phi^2 \right) + \frac{1}{2}\partial^\mu\chi\partial_\mu\chi - \frac{1}{2}\partial^\mu\phi\partial_\mu\phi - \frac{\lambda}{4}\phi^2\left( \chi-\phi \right)^2 \right] \, .
\end{equation}
The negative sign of $\chi$ field seems dangerous, but it will not bring any classical instability since it is not a physical degree of freedom.
It is invariant under the following local gauge transformations:
\begin{equation}
\label{xform}
\begin{split}
 g_{\mu\nu} & \rightarrow e^{-2\sigma(x)}g_{\mu\nu} \, ,
 \\
 \chi & \rightarrow e^{\sigma(x)}\chi \, ,
 \\
 \phi & \rightarrow e^{\sigma(x)}\phi \, .
\end{split}
\end{equation}
This gauge invariance tells us that we can fix the gauge by choosing an appropriate field $\sigma(x)$ and get the physical degree of freedom. It looks like a conformal symmetry under this gauge transformation, but it is not as the metric is not changed physically but only redefined~\cite{Hertzberg:2014aha}.
One can eliminate this degree of freedom by fixing a specific gauge, i.e. by choosing a specific function $\sigma(x)$ which may be arbitrary. A convenient choice is
\begin{equation}\label{gaugefix}
 \chi^2-\phi^2=6 \, .
\end{equation}
To realize this gauge fixing, it is also convenient to define a new field $\varphi$ in terms of which $\chi$ and $\phi$ are written as
\begin{equation}\label{NewField}
\begin{split}
 \chi & = \sqrt{6}\cosh \left( \frac{\varphi}{\sqrt{6}} \right) \, ,
 \\
 \phi & = \sqrt{6}\sinh \left( \frac{\varphi}{\sqrt{6}} \right) \, ,
\end{split}
\end{equation}
and correspondingly, the Lagrangian can be reformulated as
\begin{equation}\label{StarobinskyLag}
\calL = \sqrt{-g} \left[ \frac{R}{2} - \frac{1}{2}\partial^\mu\varphi\partial_\mu\varphi - \frac{9}{4}\lambda e^{-4\varphi/\sqrt{6}} \left( 1-e^{2\varphi/\sqrt{6}} \right)^2 \right] \,.
\end{equation}
One can easily find this Lagrangian is exactly the form of the Starobinsky model $R+\alpha R^2$ in the Einstein frame with $\alpha=(18\lambda)^{-1}$~\cite{DeFelice:2010aj}. Note that under the gauge condition (\ref{gaugefix}), slow-roll inflation occurs at $\phi/\chi\to1$, where an $SO(1,1)$ symmetry in field space is restored. 

The extension to this description, the so-called T-models~\cite{Kallosh:2013hoa}, has a general potential as compared to 
(\ref{conformalStar}),
\begin{equation}\label{conformalT}
\calL = \sqrt{-g}
\left[ \frac{R}{12} \left( \chi^2-\phi^2 \right) + \frac{1}{2}\partial^\mu\chi\partial_\mu\chi - \frac{1}{2}\partial^\mu\phi\partial_\mu\phi - \frac{1}{36} F\left(\phi/\chi\right)\left( \chi^2-\phi^2\right)^2 \right] \, ,
\end{equation}
where $F$ is a function of $\phi/\chi$.
We can see that except for this $F$ factor, all the other terms preserves an extra $SO(1,1)$ symmetry in the 
field
space of $\chi$ and $\phi$. Inflation happens in the neighborhood of the limit $\phi/\chi\rightarrow1$, where the $SO(1,1)$ symmetry is restored. As inflation 
proceeds,
$\phi/\chi$ deviates from 1 monotonously. And finally inflation ends as the $SO(1,1)$ symmetry is completely broken.
To study this deviation, we define a new variable
\begin{equation}\label{def:z}
z \equiv \frac{\phi}{\chi} = \tanh \left( \frac{\varphi}{\sqrt{6}} \right) \, .
\end{equation}
One can see from this definition that unless $\chi$ and $\phi$ are complex fields with different phases, a real $\varphi$ satisfying (\ref{gaugefix}) will generate $z<1$. A typical T-model chooses a specific form for the function $F$ as $F(z)=\lambda z^{2p}$. With the gauge condition (\ref{gaugefix}), the potential becomes simply
\begin{equation}
V(z) = F(z) \, .
\end{equation}
The predictions for $r$ and $n_s$ are very close to that of the Starobinsky model regardless the specific value of the exponent $p$~\cite{Kallosh:2013hoa}. That is, both the original Starobinsky model and the T-models predict small tensor modes as is in tension with the recent BICEP2 data.
Here we see that the $SO(1,1)$ symmetry plays a crucial rule in constructing the potential. If we preserve this symmetry in inflation, all the models generated from (\ref{conformalT}) have similar predictions. However, as we will see in the next section, there exist other generalized models originating from this Lagrangian 
that can give rise to different relations between $r$ and $n_s$.

\section{Generalized models}
\label{sec:general}

In this section we construct a broader class of 
models beyond T-models. To do this we abandon the preservation of the $SO(1,1)$ symmetry of the 
field
space even during the inflationary period, but still keep the gauge invariance (\ref{xform}). 
In this case, the general potential of the action \eqref{conformalStar} is
\begin{equation}
V(\phi,\chi) = V_0 \chi^4f(\phi/\chi) \, .
\end{equation}
%
The $\chi^4$ factor is necessary to keep the gauge invariance of the entire Lagrangian. 

After we choose the conformal gauge as (\ref{gaugefix}), this potential can be expressed solely by $z=\phi/\chi$ defined in (\ref{def:z}) as
\begin{equation}\label{V(z)}
V(z) = 36 V_0 \frac{ f(z)}{(1-z^2)^2} \, .
\end{equation}
We can see that if $f(z)$ is a smooth function, there exist two second-order poles, $z=\pm1$. In the present work, we focus our interest on the case of real fields and hence the pole at $z=-1$ will not be addressed. The existence of the pole at $z=1$, however, would probably spoil the slow-roll requirements of inflationary cosmology due to the singular mathematical behavior around this position. In order to describe the dynamics of the inflationary phase, it is convenient to define the slow-roll parameters $\epsilon$ and $\eta$ as
\begin{equation}\label{def:epsilon}
\epsilon \equiv \frac{1}{2} \left( \frac{V_\varphi}{V} \right)^2 \, ,
\end{equation}
with
\begin{equation}\label{Vslope}
\frac{V_\varphi}{V} = \frac{1}{\sqrt{6}} \left[ 4z + (1-z^2) \frac{f'}{f} \right]
\equiv \frac{g(z)}{\sqrt{6}} \, ,
\end{equation}
and
\begin{equation}\label{def:eta}
\eta \equiv \frac{\dot\epsilon}{H\epsilon}
= -\frac{1-z^2}{3}g'(z) \, ,
\end{equation}
of which the values are required to be much less than unity under the slow-roll condition. However, one can easily observe that when $z$ is close to $1$, $V_\varphi/V$ and in turn $\epsilon$ would be in general $\calO(1)$ and the slow-roll condition is not guaranteed, unless $f'/f$ contains a first order pole at $z=1$ with a residue equal to (or close to) $2$. A simple case is the T-models where $f(z)=z^{2p}(1-z^2)^2/36$~\cite{Kallosh:2013hoa}. Then both second order poles in (\ref{V(z)}) are canceled and the potential is simply $V(z) = z^{2p}$. Another important case is the original Starobinsky model~\cite{Starobinsky:1980te}. The effective potential in (\ref{StarobinskyLag}) can be recovered from (\ref{V(z)}) once we choose $f(z)=(1-z)^2/9$.

\subsection{Starobinsky-like model with dynamical exponent}

Our next step is to seek for more general models from the above requirements.
The exact form of $g(\xi)$ with $\xi \equiv 1-z$ is not clear except for its asymptotic behavior $\lim_{\xi\to0}|g(\xi)|\ll1$ to preserve the slow-roll conditions.  A non-zero limit of $\lim_{\xi\to0}g(\xi)=\Delta$ will give rise to a Starobinsky-like model equivalent to $R+\alpha R^n$ with an arbitrary exponent
$n=(2-\Delta)/(1-\Delta)$. As the predictions in this case are disfavoured by the BICEP2 result as well, we will not discuss it in detail. For further discussions, see for instance \cite{Costa:2014lta}. 
For simplicity, let us just suppose that $\lim_{\xi\to0}g(\xi)=0$ in this paper.
Even so, it is still difficult to determine the relation between $\epsilon$ and $\eta$ accurately. However, if $g'(\xi)$ is finite, we can make a Tayler expansion of $g(\xi)$ around $\xi\to0$ as
\begin{equation}\label{gexpansion}
g(\xi) = g'(0)\xi + \frac{1}{2}g''(0)\xi^2 + \cdots \, .
\end{equation}
This yields
\begin{equation}\label{eta^2}
\epsilon \approx \frac{3}{16}\eta^2
\end{equation}
up to the leading order of $\xi$. After substituting it back into (\ref{Vslope}), we can see that the Starobinsky model is a special case when $g(\xi)=2\xi$ thus $f(z)\propto(1-z)^2$.
If $g'(0)=0$, we have to go to higher order in (\ref{gexpansion}). Let us assume that the first $n-1$ terms of the Taylor series are all zero, and (\ref{gexpansion}) begins from
\begin{equation}
g(\xi)=g^{(n)}(0)\xi^n+\cdots \, .
\end{equation}
Then we find a solution
\begin{equation}
f(z)=(z-1)^2\exp\left[-\frac{g^{(n)}(0)(z-1)^n}{2n}\right] \, ,
\end{equation}
which deviates from the Starobinsky model only up to next-to-leading order.

An interesting case occurs when $g(\xi)$ is not an analytic function at $\xi=0$ and in the meanwhile $g'(\xi)$ is divergent. Namely, we consider a toy model
\begin{equation}\label{g-log}
g(\xi)=-2\lambda\xi\log\xi \, ,
\end{equation}
where $\xi=0$ is a removable singularity and $\lambda$ is a positive parameter. The minus sign is to make sure that $g(\xi)>0$ for small $\xi$.
This can still preserve the slow-roll conditions, but now we cannot have a general relationship as (\ref{eta^2}).
A particular solution to (\ref{Vslope}), up to leading order, is now given by
\begin{equation}\label{f(z)}
f(z)=(1-z)^{2+\lambda(1-z)}e^{-\lambda(1-z)} \, .
\end{equation}
This gives us an effective potential $V(\chi,\phi) \sim \chi^4(\phi/\chi-1)^{2+\lambda(\phi/\chi-1)}$, a dynamical modification to the power index.

With this modification one can calculate the slow-roll parameter $\epsilon$ as
\begin{equation}
\epsilon = \frac{\lambda^2}{3}\xi^2\left(\log\xi\right)^2 \, ,
\end{equation}
whose solution can be expressed by the Lambert $W$ function,
\begin{equation}\label{xi-e1}
\xi = -\frac{\sqrt{3\epsilon}}{\lambda W_{-1}\left(-\sqrt{3\epsilon}/\lambda\right)}
= \exp\left[W_{-1}\left(-\frac{\sqrt{3\epsilon}}{\lambda}\right)\right] \, .
\end{equation}
Note that the above two different expressions are mathematically equivalent due to the definition of the Lambert $W$ function. The subscript $-1$ denotes the lower branch of the Lambert function, where we have
\begin{equation}
0 < \frac{\sqrt{3\epsilon}}{\lambda} < \frac{1}{e}
\quad \text{and} \quad
\xi < \frac{1}{e} \, .
\end{equation}
%
Therefore, we can obtain the spectral index of the power spectrum $n_s$ in terms of the tensor-to-scalor ratio $r$ through the following relation:
\begin{equation}
n_s = 1 - \sqrt{\frac{r}{3}} \left[1+\frac{1}{W_{-1}\left(-\sqrt{3r}/\lambda\right)} \right]
\left[ 1-\frac{\sqrt{3r}}{8\lambda W_{-1}\left(-\sqrt{3r}/\lambda \right)} \right] - \frac{r}{8} \, .
\end{equation}
It is easy to check that the traditional result obtained by the Starobinsky model can be recovered under the limit $\lambda\to\infty$: $n_s \approx 1-\sqrt{r/3}$.

\begin{figure}[ht]
 \begin{center}
  \includegraphics[width=8.27cm]{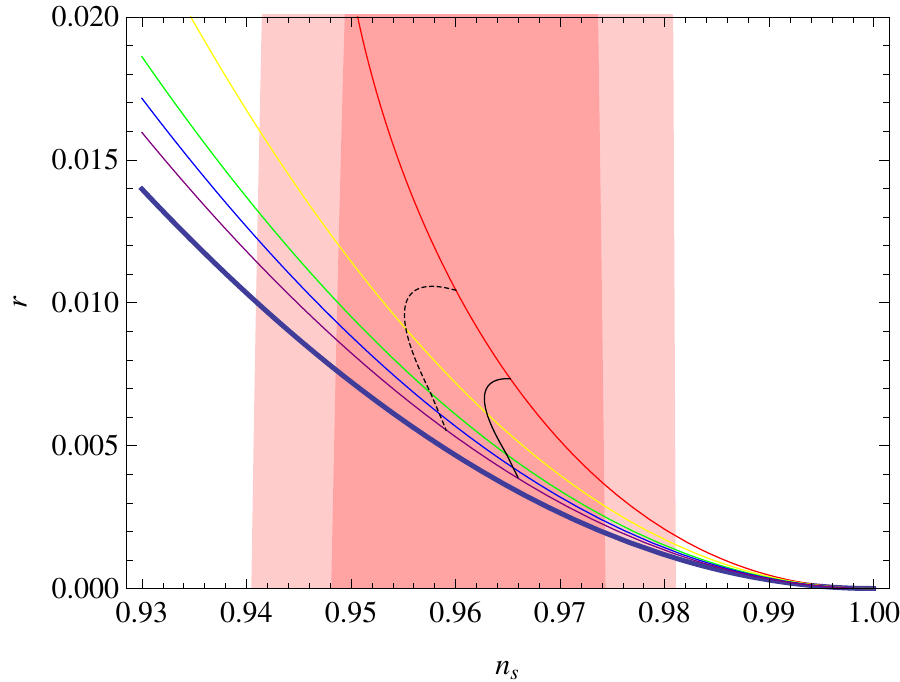}
  \hspace{1em}
  \includegraphics[width=8cm]{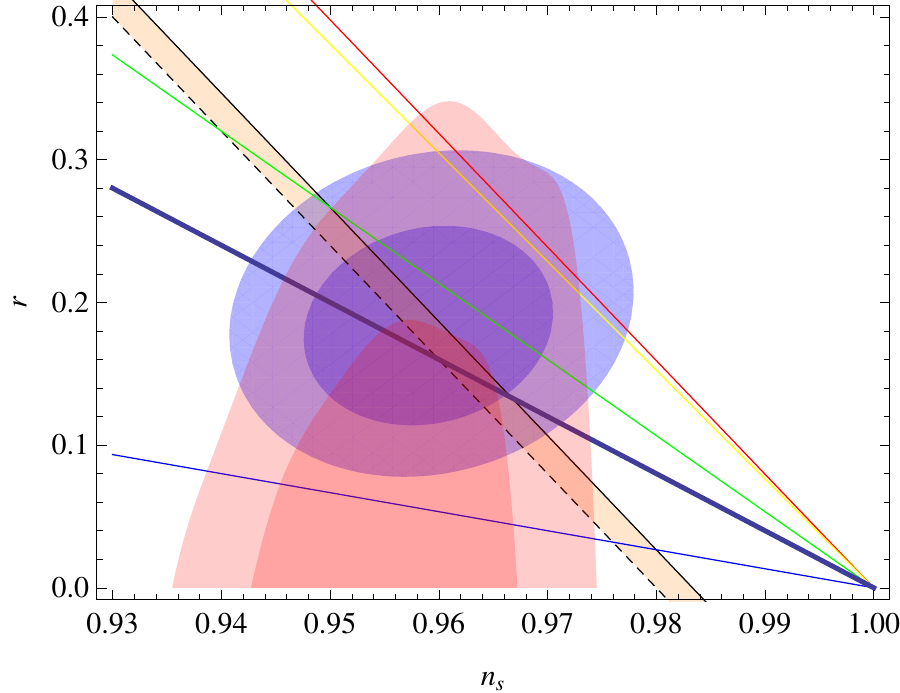}
 \end{center}
 \caption{(Left) $r$-$n_s$ diagram for (\ref{g-log}). The Planck constraints~\cite{Ade:2013uln} are shown, but the BICEP2 ones are out of range. The thick line denotes the Starobinsky model which corresponds to $\lambda\to\infty$. The thin curves are for different values of $\lambda$: from the top, $\lambda=0.25$, 1, 10, $10^2$ and $10^4$, respectively. The thin dotted and solid lines represents the $e$-folds respectively $N=50$ and 60. (Right) $r$-$n_s$ diagram for (\ref{g(z)}). The blue contours are the recent BICEP2 result. The diagonal lines correspond to different values of $\mu$: from the top, $\mu=0.01$, 0.1, 1, 2 (thick), and 10, respectively. The thin dotted and solid lines represents the $e$-folds respectively $N=50$ and 60.}
 \label{fig:2ex}
\end{figure}

The $e$-folding dependence can be found easily as
\begin{equation}\label{def:Ne}
N = \int \frac{d\varphi}{\sqrt{2\epsilon}}
= \frac{3}{\lambda} \int \frac{d\xi}{\xi^2(2+\xi)|\log\xi|}
\approx \frac{3}{2\lambda} \mathrm{li}\left(\frac{1}{\xi}\right) \, ,
\end{equation}
where li$(x)$ denotes the logarithmic integral function. We can solve the inverse relation and derive
\begin{equation}
\xi = \left[ \mathrm{li}^{(-1)} \left(\frac{2\lambda N}{3}\right) \right]^{-1} \, .
\end{equation}
This can be taken back into (\ref{xi-e1}) to have a set of parametric equations of $r$ and $n_s$ depending on $\lambda$. We show this relation in the left panel of Figure~\ref{fig:2ex}. From this plot, one can see that the value of the tensor-to-scalar ratio is in general very tiny unless one finely tune the parameter $\lambda$ to be pretty small.

\subsection{Chaotic inflation embedded in conformal description}

Now we explore another interesting extension of the conformally invariant inflation models, which is particularly of recent observable interest. As mentioned in Section~\ref{sec:intro}, the recent BICEP2 result indicates a large tensor-to-scalar ration $r$. Thus, it is an interesting question whether one can extend the conformal description to include the models which lead to a large value of $r$ consistent with the BICEP2 observation. To be clear, we would like to reconstruct the form of $g(z)$ by assuming
\begin{equation}
\eta = \mu\epsilon \, ,
\end{equation}
where $\mu$ is a dimensionless parameter of $\calO(1)$. In this case, we can solve
\begin{equation}\label{g(z)}
g(z) = \frac{4}{\mu\tanh^{-1}z}
= \frac{8}{\mu \log \left(\frac{\displaystyle 1+z}{\displaystyle 1-z} \right)} \, .
\end{equation}
Substituting it back into (\ref{Vslope}), we then obtain
\begin{equation}
f(z) = \left( 1-z^2 \right)^2 \left[ \log \left( \frac{1+z}{1-z}\right) \right]^{4/\mu} \, ,
\end{equation}
and further derive the effective potential for the inflaton field as
\begin{equation}\label{chaoticV}
V(\varphi) = V_0~ 36^{1-1/\mu}\varphi^{4/\mu} \, .
\end{equation}
From this expression, one can clearly find that this model is equivalent to chaotic inflation with a power-law potential of $\varphi$. In this regard, we point out that chaotic inflation can also be described by the conformally invariant theories under specific choices of the function $f$, or $g$ equivalently.

The number of $e$-foldings for inflation can be expressed by
%
%
%
\begin{equation}\label{N(z)}
N = \frac{3}{16} \mu \log^2 \left( \frac{1+z}{1-z} \right) \, ,
\end{equation}
%
where we have chosen the number of $e$-foldings to be vanishing when $z=0$. Note that, in a realistic model this point corresponds to the true vacuum of the inflaton field after inflation and hence, there are several number of $e$-foldings overestimated in our expression (which is associated with the preheating phase). However, this approximation would not alter the precise result and can greatly simplify the analytic analysis in the following.
By solving (\ref{N(z)}) inversely, one can easily derive the variable $z$ as a function of $N$:
\begin{equation}\label{xi-N}
z = \tanh \left( 2\sqrt{\frac{N}{3\mu}} \right) \, ,
\end{equation}
as well as the slow roll parameters
\begin{equation}
\epsilon = \frac{1}{\mu N}
\quad \text{and} \quad
\eta = \frac{1}{N} \, .
\end{equation}
This is as expected from the potential for chaotic inflation (\ref{chaoticV}). We explicitly show the predictions on the $r$-$n_s$ diagram with different values of $\mu$ in the right panel of Figure~\ref{fig:2ex}. Since the coefficient $\mu$ is a free parameter in the conformally invariant description for inflation models, we specifically consider several parameter choices by taking 
different values of $\mu$. 
One can see that the parameter choices for 
$\mu=\calO(1)$ lie
in the center of the contour provided by BICEP2. In particular, when $\mu=2$, 
which corresponds to the standard $m^2\varphi^2$ inflation model, can fit to the data very well.

\section{Conclusions}
\label{sec:conc}

As indicated by the recent BICEP2 data with a high level detection of the primordial tensor modes, it seems that a class of inflation models 
 such as T-models as well as the Starobinsky model, is disfavoured. Thus, it is interesting to study whether the formalism of the inflation models which the T-models originates remains of theoretical interest. This issue has been carried out from different viewpoints, see e.g.~
\cite{starobinsky-afterbicep2}
for relevant discussions.

In the present article, we have developed a generalized class of the T-models 
which breaks the approximate $SO(1,1)$ symmetry. We start from an arbitrary potential $V\sim f(z)\left(1-z^2\right)^{-2}$ with $z=\tanh(\varphi/\sqrt{6})$ as a dynamical field, which can describe the Starobinsky model as well as T-models as specific and straightforward examples. We interestingly find that a larger class of more general inflation models can be embedded into this formalism. {Also, a larger portion on the $r$-$n_s$ plane can be covered.} In particular, we have considered two concrete examples. One is a Starobinsky-like model which admits a dynamical exponent, which generically predicts a small tensor-to-scalar ratio $r$ and hence is 
in tension with
the recent BICEP2 observation. In the other case, we have found that chaotic inflation model can also be embedded into the conformally invariant description. In this scenario, the power index of the derived chaotic inflation can be a non-integer value as is determined by the model parameter appeared in the original Lagrangian.
As a result, for this type of models, there exist enough parameter space to accommodate with the latest cosmological observations.

\subsection*{Acknowledgements}

We thank Frederico Arroja, Qing-Guo Huang, Takahiro Tanaka and Yingli Zhang for useful discussions. We acknowledge the workshop ``New Perspectives on Cosmology'' at the Asia Pacific Center for Theoretical Physics where part of this work was carried out.
The work of YFC is supported in part by NSERC and by funds from the Canada Research Chair Program.
JG acknowledges the Max-Planck-Gesellschaft, the Korea Ministry of Education, Science and Technology, Gyeongsangbuk-Do and Pohang City for the support of the Independent Junior Research Group at the Asia Pacific Center for Theoretical Physics. JG is also supported by a Starting Grant through the Basic Science Research Program of the National Research Foundation of Korea (2013R1A1A1006701).
SP thanks the GCOE Bilateral International Exchange Program of Kyoto University and the hospitality of the Yukawa Institute for Theoretical Physics at Kyoto University for his visit.

\end{document}